\documentclass[%
reprint,
superscriptaddress,
frontmatterverbose, 
preprintnumbers,
 amsmath,amssymb,
 aps,
 prl,
]{revtex4-1}
\allowdisplaybreaks
\usepackage{graphicx}
\usepackage{dcolumn}
\usepackage{bm}
\usepackage{mathtools}
\usepackage{times}
\usepackage{nicefrac}
\usepackage{amsmath}
\usepackage{amsfonts}
\usepackage{amssymb}
\usepackage{amsthm}
\usepackage{xcolor}

\newcolumntype{w}[1]{D{.}{.}{#1}}

\begin{document}

\title{Nuclear Charge Radii of $^{10,11}$B}%
    
\author{Bernhard Maa\ss}%
\email{bmaass@ikp.tu-darmstadt.de}
\affiliation{%
Institut f\"ur Kernphysik, TU Darmstadt, 
64289 Darmstadt, Germany
}%
\author{Thomas H\"uther}%
\affiliation{%
Institut f\"ur Kernphysik, TU Darmstadt,
64289 Darmstadt, Germany
}%
\author{Jan Krause}%
\affiliation{%
Institut f\"ur Kernphysik, TU Darmstadt,
64289 Darmstadt, Germany
}%

\author{J\"org Kr\"amer}%
\affiliation{%
Institut f\"ur Kernphysik, TU Darmstadt, 
64289 Darmstadt, Germany
}%

\author{Kristian K\"onig}%
\affiliation{%
Institut f\"ur Kernphysik, TU Darmstadt, 
64289 Darmstadt, Germany
}%
\author{Alessandro Lovato}%
\affiliation{%
Physics Division, Argonne National Laboratory, 
Lemont, IL 60439, USA
}%

\author{Peter M\"uller}%
\affiliation{%
Physics Division, Argonne National Laboratory, 
Lemont, IL 60439, USA
}%
\author{Mariusz Puchalski}
\affiliation{Faculty of Chemistry, Adam Mickiewicz University, Umultowska 89b, 61-614 Pozna{\'n}, Poland}
\author{Krzysztof Pachucki}
\affiliation{Faculty of Physics, University of Warsaw, Pasteura 5, 02-093 Warsaw, Poland}
\author{Robert Roth}%
\affiliation{%
Institut f\"ur Kernphysik, TU Darmstadt,
64289 Darmstadt, Germany
}%
\author{Rodolfo S\'{a}nchez}%
\affiliation{%
GSI Helmholtzzentrum f\"ur Schwerionenforschung, 
64291 Darmstadt, Germany
}%
\author{Felix Sommer}%
\affiliation{%
Institut f\"ur Kernphysik, TU Darmstadt, 
64289 Darmstadt, Germany
}%

\author{R. B. Wiringa}%
\affiliation{%
Physics Division, Argonne National Laboratory, 
Lemont, IL 60439, USA
}%

\author{Wilfried N\"ortersh\"auser}%
\affiliation{%
Institut f\"ur Kernphysik, TU Darmstadt, 
64289 Darmstadt, Germany
}%
\noaffiliation

\date{\today}

\begin{abstract}
The first determination of the nuclear charge radius by laser spectroscopy for a five-electron system is reported. This is achieved by combining high-accuracy \textit{ab initio} mass-shift calculations and a high-resolution measurement of the isotope shift in the $2s^2 2p\, ^2\mathrm{P}_{1/2} \rightarrow 2s^2 3s\, ^2\mathrm{S}_{1/2}$ ground state transition in boron atoms. Accuracy is increased by orders of magnitude for the stable isotopes $^{10,11}$B and the results are used to extract their difference in the mean-square charge radius $\langle r^2_\mathrm{c}\rangle^{11} - \langle r^2_\mathrm{c}\rangle^{10} = -0.49\,(12)\,\mathrm{fm}^2$. The result is qualitatively explained by a possible cluster structure of the boron nuclei and quantitatively used as a benchmark for new \textit{ab initio} nuclear structure calculations using the no-core shell model and Green's function Monte Carlo approaches.

\end{abstract}

\pacs{Valid PACS appear here}

\maketitle

\textit{Introduction}---The lightest elements play an exceptional role for the advancement of nuclear and atomic physics: Only here theoretical approaches are sufficiently advanced to calculate both electronic and the nuclear structure from first principles. Laser spectroscopy provides unique benchmarks to test and further advance those models of the fundamental structure of nature. For hydrogen-like systems, atomic theory is sufficiently accurate to calculate transition frequencies including quantum electrodynamic (QED) corrections to such a precision that the mean-square nuclear charge radius $\left\langle r_\mathrm{c}^2 \right\rangle$ can be extracted. This has been demonstrated for hydrogen~\cite{Udem1997}, muonic hydrogen~\cite{Pohl2010}, and muonic deuterium~\cite{Pohl2016}. But already for two-electron systems this is so far not feasible even though first progress towards this goal has been reported~\cite{Yerokhin2018}. However, calculating the mass-dependent isotope shift $\delta \nu_\mathrm{MS}^{A,A'}$ between two isotopes $A$ and $A'$ in an optical transition has the advantage that all mass-independent contributions and their related uncertainties cancel. This allows to isolate the small portion of the isotope shift that is caused by the change in the mean-square nuclear charge radius $\delta \left\langle r_\mathrm{c}^2 \right\rangle$ between isotopes. A few experiments have already utilized this technique to obtain nuclear charge radii of stable and short-lived isotopes of He \cite{Wang2004,Mueller2007}, Li \cite{Ewald2005,Sanchez2006}, and Be$^+$ \cite{Noertershaeuser2009,Krieger2012}, based on corresponding mass-shift calculations in two- and three-electron systems \cite{Yan2001,Yan2008,Pachucki2003,Puchalski2008}. 

Here, we report the first application of this technique to the five-electron system of atomic boron. We present high-precision calculations of the mass shift and the field-shift factor required to extract the difference in mean-square nuclear charge radius $\delta \left \langle r_\mathrm{c}^2 \right \rangle$ between the two stable boron isotopes $^{10,11}$B and a measurement of the isotope shift using resonance ionization mass spectrometry on a thermal atomic beam. The results are compared to new \textit{ab initio} nuclear structure calculations. \\
\textit{Isotope Shift Calculations}--- The approach to extract $\delta \left \langle r_\mathrm{c}^2 \right \rangle$ between $^{10}$B and $^{11}$B is similar to our previous work \cite{Lu2013}: The isotope shift is composed of the mass shift (MS) and the field shift (FS) $\delta \nu_\mathrm{IS} = \delta \nu_\mathrm{MS} + \delta \nu_\mathrm{FS}$, where the latter $\delta \nu_\mathrm{FS} = C \, \delta \left \langle r_\mathrm{c}^2 \right \rangle$ contains the information on the charge radius difference. To obtain $\delta \nu_\mathrm{MS}$ with the required high accuracy we calculate the shift due to the finite nuclear mass in powers of the fine structure constant $\alpha$. Namely, the atomic energy levels are considered as a function of $\alpha$, which is expanded in a power series 
\begin{eqnarray}
E(\alpha) &=& \sum_{n} E^{(n)}\,, \quad E ^{(n)}\,\sim m \alpha^n,\quad n=2,4,5,6,\dots\quad .
\label{Eexpansion}
\end{eqnarray}
These expansion coefficients are calculated including finite nuclear mass effects but neglecting the nuclear spin, which gives rise to a hyperfine splitting but does not shift the energy levels in first order.

The leading term $E^{(2)}$ is the eigenvalue of the non-relativistic Hamiltonian
in the center-of-mass system. This Hamiltonian is used to obtain the non-relativistic wave function $\Psi$ using the variational approach, where it is expressed in the form of $K$-term linear combinations of the five-electron basis functions
$\psi_l({\vec r})$
\begin{equation}\label{08}
\Psi({\bf r},\boldsymbol{\sigma})=
\hat{\mathcal A}\left(\Xi_{S,M_S}(\vec{\sigma})\,
     \sum_{l=1}^{K}c_l\,\psi_l({\vec r})\right).
\end{equation}
The operator $\hat{\mathcal A}$ ensures the antisymmetry of the total wave
function with respect to the exchange of the electrons. The
$\Xi_{S,M_S}(\vec{\sigma})$ is an $n$-electron spin eigenfunction
with the quantum numbers $S$ and $M_S$, and $\vec{\sigma}$ and ${\vec r} $
are the $n$-electron vectors in spin and coordinate space.  The
spatial basis functions are the five-electron explicitly correlated Gaussian (ECG) functions
of S and P symmetry, respectively:
\begin{align}
\psi_l({\vec{r}})=&\ \exp\Bigl[ -\sum_{a>b} c_{ab}\,(\vec r_a-\vec r_b)^2 \Bigr], \\
    {\vec \psi_l}({\vec{r}})=&\ {\vec r}_{a}\,\exp\Bigl[ -\sum_{a>b} c_{ab}\,(\vec r_a-\vec r_b)^2 \Bigr],
\end{align}
with ${\vec r}_{a}$ being the coordinate of the $a$-th particle (electrons and the nucleus).  The linear parameters $c_l$ are obtained by the standard inverse iteration method. The nonlinear parameters $c_{ab}$ are determined variationally for each basis function in an extensive optimization of the non-relativistic energy $E^{(2)}$ with progressively doubled size from $K=1024$ to $K=8192$ terms \cite{Puchalski2015}. 

The obtained non-relativistic wave functions are then used for further perturbative calculations of relativistic and QED contributions using the respective Hamiltonians. Finite nuclear mass terms of the non-relativistic energy are treated up to $(m/M)^2$ and those of higher order in $\alpha$ up to $m/M$. Accordingly, $E^{(4)} = \langle\Psi|H^{(4)}|\Psi\rangle$ is calculated using the relativistic Hamiltonian $H^{(4)}$ and the non-relativistic wave function $\Psi$. The similar calculation for the leading QED correction $E^{(5)} = \langle\Psi|H^{(5)}|\Psi\rangle$ can be found in the supplemental material. At present, a complete numerical evaluation of $E^{(6)}$ correction  for a five-electron system is unfeasible, as the full calculation of $E^{(6)}$ has been performed only for one- and two-electron systems \cite{Patkos2016,Patkos2017}. Based on this experience, the $E^{(6)}$ term is estimated using its dominating contribution built of the leading one-electron terms which are proportional to the contact term $\delta^3(r_a)$. No higher order terms are needed here, since the related uncertainty is much smaller than the one from $E^{(4)}$. Following the convention introduced for $E^{(n)}$, formulas for related contributions to the transition energy $\nu$ between the atomic states $X$ and $Y$ and to the mass shift between the isotopes can be written as 
\begin{eqnarray}
\nu^{(n)}(X \rightarrow Y) &=& E^{(n)}(Y) - E^{(n)}(X) \\
\delta^{(n)} \nu_\mathrm{MS}  &=& \nu^{(n)}(^{10}\mathrm{B}) - \nu^{(n)}(^{11}\mathrm{B})\,.
\end{eqnarray}
and are summarized in Tab.\,\ref{T:Efs}. To determine the field shift factor C we consider the leading $r_c$-dependent correction
\begin{align}
E_{\rm fs}^{(4)} &=  \frac{2\,\pi}{3}\,Z\,\alpha\,\langle r_\mathrm{c}^2 \rangle\,\sum_{a} \langle  \delta^3(r_{a}) \rangle
\label{E40fs}
\end{align}
and the logarithmic relativistic correction to the wave function at the origin
\begin{equation}
E_{\rm fs,log}^{(6)} = -(Z\,\alpha)^2\,\ln(Z\,\alpha\,m\,\langle r_{\rm c}^2\rangle)\,E^{(4)}_{\rm fs}\,.
\label{E60logfs}
\end{equation}
Our recommended value for the constant $C$ is obtained as a sum of two components $C = C^{(4)}+C^{(6)}_\mathrm{log}$ and an arithmetic average over the isotopes $C = (C_{11}+C_{10})/2$. The remaining term $(C_{11}-C_{10})\,(\langle r_\mathrm{c}^2 \rangle^A + \langle r_\mathrm{c}^2\rangle^{A'})/2$ is as small as the neglected relativistic $\alpha^4\,m^3/M^2$ corrections and is therefore also neglected. The result is given in Tab.\,\ref{T:Efs} with an uncertainty estimated as 50\% of the logarithmic correction. More details on this calculation can be found in the supplemental material.
\begin{table}[thb]
\renewcommand{\arraystretch}{1.3}
\caption{Components of the isotope shift $\delta \nu_\mathrm{IS} = \nu \left( ^{10}\mathrm{B}\right) - \nu \left( ^{11}\mathrm{B}\right)$ in the  $2p\,^2P_{1/2} \rightarrow 3s\,^2S_{1/2}$ transition of the boron atom.}
\label{T:Efs}
\begin{ruledtabular}
\begin{tabular}{c @{\hspace{1in}}d@{$\pm$}d}
Component    & \multicolumn{2}{c}{Value (MHz)}  \\
\hline
$m\,\alpha^2 $                           & 5027.27&0.03 \\
$m\,\alpha^4$                            &   -4.78&0.07 \\
$m\,\alpha^5$         			 &    0.572&0.005   \\
$m\,\alpha^6$       		         &   -0.058&0.002   \\[1ex]
Total                        		 & 5023.00&0.08 \\[1ex]
$C [{\rm MHz/fm^2}]$         		 &  16.91&0.09
\end{tabular}
\end{ruledtabular}
\end{table}

\begin{figure}[b]
\includegraphics[width=0.48\textwidth]{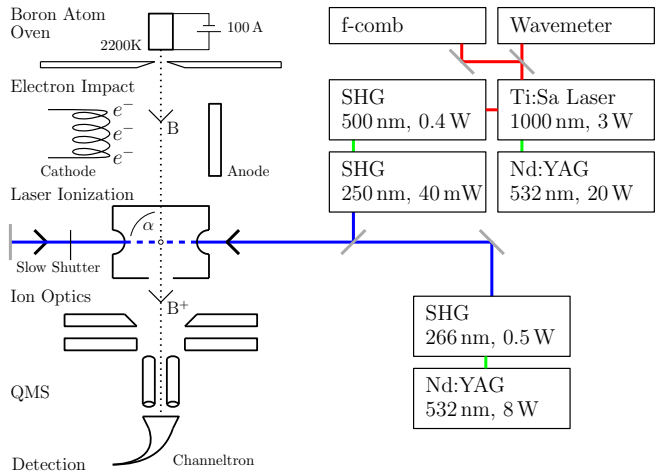}
\caption{\label{fig:setup} Experimental setup: A tunable, continuous-wave laser system is generating 250\,nm light via second harmonic generation (SHG) while being monitored with a frequency comb and a HighFinesse Wavemeter (Fizeau interferometer). The laser beam is superimposed perpendicular with a thermal atomic beam of boron and back-reflected for a second pass through the interaction region. A second laser system is used for non-resonant ionization of the excited atoms. Ion detection after mass selection in a quadrupole mass spectrometer (QMS) is performed with a Channeltron detector.}
\end{figure}

\textit{Experiment}---Resonance ionization spectroscopy was performed in the two-step ionization scheme
\begin{equation}
2s^2\,2p\,^2\mathrm{P}_{1/2} \xrightarrow[\text{249.75\,nm}]{\lambda_1} 
2s^2\,3s\,^2\mathrm{S}_{1/2} \xrightarrow[<\text{271\,nm}]{\lambda_2}
\mathrm{B^+ + e^-}
\end{equation} 
of neutral boron on a thermal atomic beam. The experimental setup is shown in Fig.\,\ref{fig:setup}. Stable boron atoms are emitted from a graphite tube filled with amorphous boron powder which is heated to approximately 2500\,K. The beam traverses first an electron-impact ionization segment used to monitor the boron production rate in between the laser spectroscopic measurements and then enters the laser interaction region. The apertures along this path limit the angular spread of the atomic beam to 2\,mrad. 

For laser excitation and ionization two single-mode continuous-wave laser systems are used: A titanium-sapphire (Ti:Sa) laser is generating light output at approximately 1000\,nm and is frequency-doubled twice to generate the resonance transition wavelength of $\lambda_1 = 249.75$\,nm. The Ti:Sa frequency is scanned stepwise during the experiment, and stabilized against short term fluctuations via an optical reference cavity at each measurement point. A frequency comb is employed to monitor long-term drifts of the reference cavity. The output power of the resonant laser is kept below 1\,mW to avoid power broadening and to limit ionization by the resonant laser light. For non-resonant ionization, a commercial cw frequency-doubled Nd:YAG Laser at 532\,nm is externally frequency-doubled to $\lambda_2=266$\,nm providing a minimum of 500\,mW power. The resonant and the non-resonant laser beam are overlapped before being intersected perpendicularly with the boron beam. 

At resonance, the tunable laser excites boron atoms to the $3s$ state from where they are ionized by absorption of another photon from either the resonant laser or the  ionization laser. The generated ions are guided electrostatically into a quadrupole mass spectrometer (QMS) for mass separation and single-ion detection at low background using a channeltron detector. The data acquisition system records the number of detected ions as a function of the frequency of the resonant laser.  
After the combined laser beams intersect the atomic beam once, they are back-reflected from a mirror outside the vacuum chamber and overlapped with a precision better than 1\,mrad to the incoming laser beams. 

To obtain a single spectrum, the laser is scanned and stabilized in steps of 8 to 10\,MHz covering the atomic resonance which spans 1.6\,GHz and 2.4\,GHz for $^{10}$B and $^{11}$B, respectively. For each frequency step a measurement with (DP = double pass) and without (SP = single pass) retroreflected beam is taken by switching a remote-controlled shutter placed in front of the reflecting mirror. The dwell time in each spectrum is adapted between 1 and 10\,s according to ionization laser power, resonant laser power and boron source performance so that a count of more than 200 photoions per step is achieved at a peak. After recording a resonance signal for $^{11}$B, which has 80\% natural abundance, laser and QMS are tuned to settings for $^{10}$B and a spectrum of this isotope is recorded under identical experimental conditions but with increased dwell time in SP configuration to improve statistics at the lower abundance. The intersection angle $\alpha$ of atom and laser beam is then deliberately varied around $\pi/2$ before another set of spectra is recorded. 

\begin{figure}[b]
\includegraphics[width=0.5\textwidth]{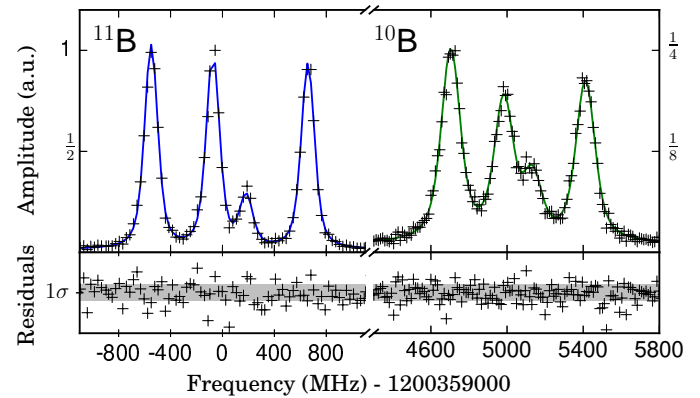}
\caption{\label{fig:full_spectra} Fitted single-pass hyperfine spectra of $^{11}$B (left) and $^{10}$B (right).}
\end{figure}
Single-pass spectra are fitted using Voigt profiles for the individual hyperfine components and shared widths for all hyperfine components. We take into account the natural Lorentzian linewidth of $40.1$\,MHz and Gaussian broadening effects originating from the thermal atom beam. In the $^{11}$B DP spectra, the transition is probed and recorded twice with a small shift in frequency $\Delta \nu_\mathrm{DP}^{11}$ since the intersection angle has never been exactly $\pi/2$ and thus the rest-frame transition frequency for the two laser beam orientations is Doppler-shifted in opposite directions. To reduce the number of degrees of freedom in the fit with back-reflection, the hyperfine factors are fixed to their literature value for the $2p$ state and the values extracted from a single-pass fit for the $3s$ level. It is assumed that the peaks from both directions each share the same width but have different intensity due to power losses and differences in beam diameter after retro-reflection. Since the single-pass spectrum was recorded simultaneously using the shutter system, one of the transition centroids for the double-pass spectra was fixed to the value from the single-pass. With these constraints, the model for the double-pass spectra converged with good agreement. An example dataset with the respective fit results is shown in Fig.\,\ref{fig:full_spectra}. 
The spectra of $^{11}$B and $^{10}$B are taken under the same ambient and spatial conditions, the transition frequency centroids are nevertheless Doppler-shifted with different magnitudes due to their difference in the mass-dependent mean velocity. The size of this effect can be determined from the double-pass spectrum of $^{11}$B and scaled for the thermal velocities: since the atoms are emitted from a thermal source, both isotopes inherit the same average kinetic energy, resulting in a velocity ratio of $r = v(^{11}\mathrm{B})/v(^{10}\mathrm{B}) = \sqrt{m(^{10}\mathrm{B})/m(^{11}\mathrm{B})}.$
The isotope shift between the stable isotopes can therefore be calculated as  
\begin{eqnarray}
\delta \nu_\mathrm{IS} = \nu_\mathrm{SP}(^{10}\mathrm{B}) - \nu_\mathrm{SP}(^{11}\mathrm{B}) - \Delta \nu_\mathrm{DP}(^{11}\mathrm{B})/2 \cdot r
\label{eq:is_correlated}
\end{eqnarray}
where $\nu_\mathrm{SP}(^{A}\mathrm{B})$ is the respective Doppler-shifted transition centroid from the single-pass spectra of $^{11}$B and $^{10}$B. With this approach, the determination of the isotope shift depends mainly on the parameters extracted from  single-pass spectra, which are determined with much smaller systematic and statistical uncertainties. 

\begin{figure}
\includegraphics[width=0.48\textwidth]{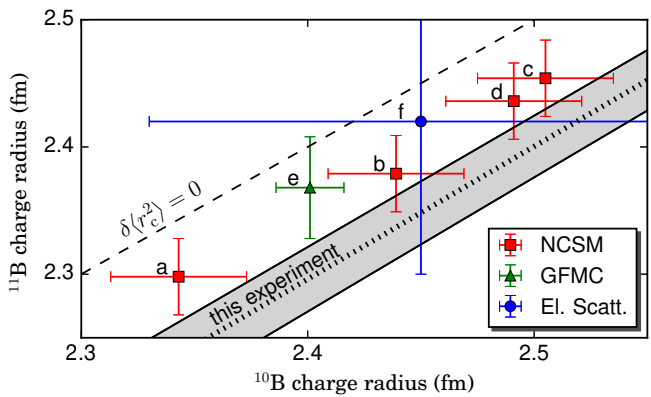}
\caption{\label{fig:rms_comp} Nuclear charge radii as obtained from various nuclear structure calculations. Results from no-core shell model calculations with different nuclear potentials are plotted as squares. The result from a Green's function Monte Carlo calculation is depicted as a triangle. For comparison, experimental data from an electron scattering experiment \cite{Stovall1966} is plotted as a circle. The letters refer to the explanation in the text and Tab.\,SI in the supplemental material. Our experimental determination of $\delta \langle r_\mathrm{c}^2 \rangle$ restricts the charge radii to lie along the dotted line within the grey uncertainty band.}
\end{figure}

Combining all data yields a value for the isotope shift of -5031.3\,(2.0)\,MHz. Table \ref{tab:is_shift} lists previous measurements of the isotope shift as well as our theoretical and experimental results. None of the previous publications denotes the isotope shift explicitly but report transition frequencies of the two isotopes with respective uncertainties. The influence of correlated uncertainties in the calculation of $\delta \nu_\mathrm{IS}$ is therefore unknown and Gaussian error propagation is used. Our value agrees within $1.3\sigma$ with the most recent value obtained by Johansson \cite{Johansson1993} but has about two orders of magnitude higher accuracy. It is close to the calculated mass shift of $-5023.00(8)$\,MHz and the difference of  $8.8\,(2.0)$\,MHz can be attributed to the finite size effect. This corresponds to a change in mean-square nuclear charge radius of $\delta \langle r_\mathrm{c}^2 \rangle = -0.49\,(12)\,$fm$^2$ taking the calculated field shift factor $C= 16.91\,(9)\,$MHz/fm$^2$ into account. At first glance it might seem counterintuitive that the charge radius of $^{10}$B is larger than that in $^{11}$B, but it is well in line with similar results in the lithium and beryllium chains \cite{Ewald2005,Sanchez2006,Noertershaeuser2009,Krieger2012}. There, a minimum of the charge radius has been observed at $N=6$ and was attributed to the cluster structure of the lighter nuclei. Similar arguments can also be used for boron: $^{9}$B is unbound since it consists of two $\alpha$-clusters and an additional proton that does not support binding of the two $\alpha$'s like the neutron does in the case of $^{9}$Be. $^{10}$B can be considered to be of $\alpha + \alpha + d$ structure and is expected to have a rather large charge radius and the decrease towards $^{11}$B would be analog to the decrease of $\delta \langle r_\mathrm{c}^2 \rangle = -0.734\,(40)\,$fm$^2$ from $^{6}$Li ($\alpha + d$) to $^{7}$Li ($\alpha + t$) \cite{Noertershaeuser2011}, which is indeed of similar size. In Fig.\,\ref{fig:rms_comp} the charge radius of $^{11}$B is plotted versus the charge radius of $^{10}$B. Apart from these more empirical arguments we compare the extracted change of the charge radius to new \textit{ab initio} calculations of the stable boron isotopes in the following.

\begin{table}
\caption{\label{tab:is_shift}%
Isotope shift $\delta \nu_\mathrm{IS} = \nu \left( ^{10}\mathrm{B}\right) - \nu \left( ^{11}\mathrm{B}\right)$  in the $2p\,^2\mathrm{P}_{1/2}\rightarrow 3s \, ^2 \mathrm{S}_{1/2}$ transition of atomic boron. The calculated mass shift listed in Tab.\,\ref{T:Efs} is used to extract the field shift that is related to the finite nuclear size.
}
\begin{ruledtabular}
\begin{tabular}{ld@{$\pm$}dl}
\textrm{}&
\multicolumn{1}{c}{\textrm{Value (MHz)}}&&\textrm{Reference}\\
\colrule
Isotope Shift &  5250&360 & \cite{Mrozowski1939}\\
 &  4110&360 & \cite{Burke1955}\\
 &  5220&150 & \cite{Johansson1993}\\
 &  5031.3&2.0 & this work, Exp. \\ 
Mass Shift &  5023.00&0.08 & this work, Theory\\
Field Shift &  8.8&2.0 & extracted\\
\end{tabular}
\end{ruledtabular}
\end{table}
\textit{Nuclear Structure Theory}---We employ two state-of-the-art \textit{ab initio} nuclear structure methods, the no-core shell model (NCSM) and the Green's function Monte Carlo (GFMC) approach, to compute the charge radii of $^{10}$B and $^{11}$B. 

For the NCSM we use different two-nucleon (NN) and three-nucleon (3N) interactions from chiral effective field theory: (a) the N2LO-SAT interaction with NN and 3N interaction at N2LO with flow parameter $\alpha=0.08\,\text{fm}^4$ \cite{Ekstroem2015}; (b) the NN interaction at N3LO by Entem and Machleidt \cite{Entem2003} supplemented with a 3N interaction at N2LO with local regulator and reduced cutoff ($\Lambda_{\text{3N}}=400\,\text{MeV/c}$, $\alpha=0.08\,\text{fm}^4$) that has been widely used in the past years \cite{Roth2012,Hergert2013,Gebrerufael2017}; (c) the Entem-Machleidt NN interaction at N3LO with a new 3N interaction at N2LO with nonlocal regulator ($\Lambda=500\,\text{MeV/c}$, $c_D=0.8$, $\alpha=0.12\,\text{fm}^4$); and (d) the recent NN interaction at N4LO by Entem, Machleidt, and Nosyk \cite{Entem2017} plus a 3N interaction at N2LO with nonlocal regulator ($\Lambda=500\,\text{MeV/c}$, $c_D=-1.8$, $\alpha=0.16\,\text{fm}^4$) \cite{Huether}. Only (a) uses information beyond the few-body sector and explicit constraints on nuclear radii to determine the low-energy constants, in all other cases the NN interaction is fitted exclusively to two-nucleon scattering data and the 3N interaction to the triton binding energy, the triton $\beta$-decay half-life, or properties of the $\alpha$ particle. For all interactions we employ a consistent similarity renormalization group evolution up to the three-body level for the Hamiltonian and up to the two-body level for the radius operator. We have confirmed that the impact of variations of the flow parameter $\alpha$ on the radii is much smaller than the model-space convergence uncertainties. For each interaction large-scale NCSM calculations are performed for model spaces from $N_{\max}=2$ to $10$ using harmonic oscillator frequencies $\hbar\Omega=12,13,...,18$ MeV. To extract the nominal value and uncertainty for the point-proton radius, we first identify the $\hbar\Omega$-value that provides the most stable radius as function of $N_{\max}$ and then use the neighboring $\hbar\Omega$-values and the residual $N_{\max}$-dependence to estimate the many-body uncertainties. 

Green’s function Monte Carlo (GFMC) uses imaginary-time projection techniques to solve for nuclear ground- and excited-state energies at the 1\% accuracy level for a given Hamiltonian~\cite{Carlson2015}. Here we employ the AV18+IL7 Hamiltonian, containing the Argonne $v_{18}$ NN
potential~\cite{Wiringa1995} and Illinois-7 3N potential~\cite{Pieper2001,Pieper2008}.
The four parameters characterizing this 3N potential were fit to low-lying nuclear spectra in the mass range A=3-10 and give an excellent reproduction of approximately 100 ground- and excited-state energies up to A=12. GFMC calculations successfully predicted the charge radii of $^{6,8}$He isotopes, while Li and Be isotope radii tend to be a little smaller than current experimental values~\cite{Lu2013}.

While our calculated binding energies for $^{10}$B agrees well with the experimental value, we slightly underbind $^{11}$B. To remedy this shortcoming of the AV18+IL7 interaction, when computing the radius of $^{11}$B we slightly quench the phenomenological repulsive term of the 3N force to reproduce the experimental binding energy.

The results for charge radii and ground state energies from both NCSM and GFMC calculations are summarized in Tab.\,SI of the supplemental material. As in Ref.~\cite{Lonardoni2018} and \cite{Lu2013}, the charge radii are derived from the estimates for the point-proton radii taking into account the finite-size of the nucleons and the Darwin-Foldy correction. The one-body spin-orbit correction of Ref.~\cite{Lu2013} has been estimated in a variational Monte Carlo calculation; it is found to be about five times smaller than the Darwin-Foldy term in $^{11}$B, and to vanish in $^{10}$B, so it is neglected here.  Two-body terms in the charge operator have also been neglected in this analysis, as their contribution was proven to be small in the GFMC calculations of the $^{12}$C charge form factor \cite{Lovato2013}.

\textit{Conclusion and Outlook}---
In Fig.\,\ref{fig:rms_comp} the charge radius of $^{11}$B is plotted versus that of $^{10}$B. 
Our newly obtained value for the difference in mean-square charge radius $\delta \langle r_\mathrm{c}^2 \rangle =  -0.49\,(12)\,\mathrm{fm}^2$ is depicted as dotted line with grey shaded uncertainty area. It improves the experimental precision from electron scattering (f) by at least a factor of five. Our value is used as benchmark for \textit{ab initio} nuclear structure theory. Independent of the applied many-body method and interaction, both NCSM (a-d) and GFMC (e) predict values which are slightly smaller, but are consistent with the experimental value for $\delta \langle r_c^2 \rangle$ within uncertainties. Absolute values of the radii vary with different interactions, which is partly correlated to a variation of the ground-state energies of the two isotopes; for the NCSM calculations with interaction (a) both isotopes are overbound by about 4 MeV compared to experiment, while for interactions (b,c,d) they are all underbound by about 4 MeV. For GFMC the ground-state energies are reproduced by construction and tweaking the 3N potential to match the experimental binding energies. 

In summary, the value obtained from high precision laser spectroscopy and novel atomic theory calculations for the 5e$^-$ system show good agreement with \textit{ab initio} nuclear structure calculations. The advances made in all three fields of physics open the window to explore a wider area of light nuclei to be investigated with high precision experiments that will provide valuable benchmarks for understanding and refining our nuclear models. \\
\textit{Acknowledgement}---We wish to acknowledge our special thanks to the late Steven C. Pieper, who contributed so much to the progress of nuclear quantum Monte Carlo. 

This work is funded by the Deutsche Forschungsgemeinschaft (DFG, German Research Foundation) - Projektnummer 279384907 - SFB 1245 and by the U.S. DOE, Office of Science, Office of Nuclear Physics, under contract DE-AC02-06CH11357.
MP and KP acknowledge support from National Science Center (Poland) Grant No. 2014/15/B/ST4/05022. 
FS and BM acknowledge support from HGS-HIRe.
Part of the NCSM calculations were performed on the LICHTENBERG high-performance computer at TU Darmstadt.
Under an award of computer time provided by the INCITE program, this research used resources of the Argonne Leadership Computing Facility at Argonne National Laboratory. 

\bibliography{quellen}

\providecommand{\noopsort}[1]{}\providecommand{\singleletter}[1]{#1}%
\begin{thebibliography}{36}%
\makeatletter
\providecommand \@ifxundefined [1]{%
 \@ifx{#1\undefined}
}%
\providecommand \@ifnum [1]{%
 \ifnum #1\expandafter \@firstoftwo
 \else \expandafter \@secondoftwo
 \fi
}%
\providecommand \@ifx [1]{%
 \ifx #1\expandafter \@firstoftwo
 \else \expandafter \@secondoftwo
 \fi
}%
\providecommand \natexlab [1]{#1}%
\providecommand \enquote  [1]{``#1''}%
\providecommand \bibnamefont  [1]{#1}%
\providecommand \bibfnamefont [1]{#1}%
\providecommand \citenamefont [1]{#1}%
\providecommand \href@noop [0]{\@secondoftwo}%
\providecommand \href [0]{\begingroup \@sanitize@url \@href}%
\providecommand \@href[1]{\@@startlink{#1}\@@href}%
\providecommand \@@href[1]{\endgroup#1\@@endlink}%
\providecommand \@sanitize@url [0]{\catcode `\\12\catcode `\$12\catcode
  `\&12\catcode `\#12\catcode `\^12\catcode `\_12\catcode `\%12\relax}%
\providecommand \@@startlink[1]{}%
\providecommand \@@endlink[0]{}%
\providecommand \url  [0]{\begingroup\@sanitize@url \@url }%
\providecommand \@url [1]{\endgroup\@href {#1}{\urlprefix }}%
\providecommand \urlprefix  [0]{URL }%
\providecommand \Eprint [0]{\href }%
\providecommand \doibase [0]{http://dx.doi.org/}%
\providecommand \selectlanguage [0]{\@gobble}%
\providecommand \bibinfo  [0]{\@secondoftwo}%
\providecommand \bibfield  [0]{\@secondoftwo}%
\providecommand \translation [1]{[#1]}%
\providecommand \BibitemOpen [0]{}%
\providecommand \bibitemStop [0]{}%
\providecommand \bibitemNoStop [0]{.\EOS\space}%
\providecommand \EOS [0]{\spacefactor3000\relax}%
\providecommand \BibitemShut  [1]{\csname bibitem#1\endcsname}%
\let\auto@bib@innerbib\@empty
\bibitem [{\citenamefont {Udem}\ \emph {et~al.}(1997)\citenamefont {Udem},
  \citenamefont {Huber}, \citenamefont {Gross}, \citenamefont {Reichert},
  \citenamefont {Prevedelli}, \citenamefont {Weitz},\ and\ \citenamefont
  {H\"ansch}}]{Udem1997}%
  \BibitemOpen
  \bibfield  {author} {\bibinfo {author} {\bibfnamefont {T.}~\bibnamefont
  {Udem}}, \bibinfo {author} {\bibfnamefont {A.}~\bibnamefont {Huber}},
  \bibinfo {author} {\bibfnamefont {B.}~\bibnamefont {Gross}}, \bibinfo
  {author} {\bibfnamefont {J.}~\bibnamefont {Reichert}}, \bibinfo {author}
  {\bibfnamefont {M.}~\bibnamefont {Prevedelli}}, \bibinfo {author}
  {\bibfnamefont {M.}~\bibnamefont {Weitz}}, \ and\ \bibinfo {author}
  {\bibfnamefont {T.~W.}\ \bibnamefont {H\"ansch}},\ }\href {\doibase
  10.1103/PhysRevLett.79.2646} {\bibfield  {journal} {\bibinfo  {journal}
  {Phys. Rev. Lett.}\ }\textbf {\bibinfo {volume} {79}},\ \bibinfo {pages}
  {2646} (\bibinfo {year} {1997})}\BibitemShut {NoStop}%
\bibitem [{\citenamefont {Pohl}\ \emph {et~al.}(2010)\citenamefont {Pohl} \emph
  {et~al.}}]{Pohl2010}%
  \BibitemOpen
  \bibfield  {author} {\bibinfo {author} {\bibfnamefont {R.}~\bibnamefont
  {Pohl}} \emph {et~al.},\ }\href {https://doi.org/10.1038/nature09250}
  {\bibfield  {journal} {\bibinfo  {journal} {Nature}\ }\textbf {\bibinfo
  {volume} {466}},\ \bibinfo {pages} {213 EP } (\bibinfo {year}
  {2010})}\BibitemShut {NoStop}%
\bibitem [{\citenamefont {Pohl}\ \emph {et~al.}(2016)\citenamefont {Pohl} \emph
  {et~al.}}]{Pohl2016}%
  \BibitemOpen
  \bibfield  {author} {\bibinfo {author} {\bibfnamefont {R.}~\bibnamefont
  {Pohl}} \emph {et~al.},\ }\href {\doibase 10.1126/science.aaf2468} {\bibfield
   {journal} {\bibinfo  {journal} {Science}\ }\textbf {\bibinfo {volume}
  {353}},\ \bibinfo {pages} {669} (\bibinfo {year} {2016})}\BibitemShut
  {NoStop}%
\bibitem [{\citenamefont {Yerokhin}\ \emph {et~al.}(2018)\citenamefont
  {Yerokhin}, \citenamefont {Patk\'o\ifmmode~\check{s}\else \v{s}\fi{}},\ and\
  \citenamefont {Pachucki}}]{Yerokhin2018}%
  \BibitemOpen
  \bibfield  {author} {\bibinfo {author} {\bibfnamefont {V.~A.}\ \bibnamefont
  {Yerokhin}}, \bibinfo {author} {\bibfnamefont {V.~c.~v.}\ \bibnamefont
  {Patk\'o\ifmmode~\check{s}\else \v{s}\fi{}}}, \ and\ \bibinfo {author}
  {\bibfnamefont {K.}~\bibnamefont {Pachucki}},\ }\href {\doibase
  10.1103/PhysRevA.98.032503} {\bibfield  {journal} {\bibinfo  {journal} {Phys.
  Rev. A}\ }\textbf {\bibinfo {volume} {98}},\ \bibinfo {pages} {032503}
  (\bibinfo {year} {2018})}\BibitemShut {NoStop}%
\bibitem [{\citenamefont {Wang}\ \emph {et~al.}(2004)\citenamefont {Wang} \emph
  {et~al.}}]{Wang2004}%
  \BibitemOpen
  \bibfield  {author} {\bibinfo {author} {\bibfnamefont {L.-B.}\ \bibnamefont
  {Wang}} \emph {et~al.},\ }\href {\doibase 10.1103/PhysRevLett.93.142501}
  {\bibfield  {journal} {\bibinfo  {journal} {Phys. Rev. Lett.}\ }\textbf
  {\bibinfo {volume} {93}},\ \bibinfo {pages} {142501} (\bibinfo {year}
  {2004})}\BibitemShut {NoStop}%
\bibitem [{\citenamefont {Mueller}\ \emph {et~al.}(2007)\citenamefont {Mueller}
  \emph {et~al.}}]{Mueller2007}%
  \BibitemOpen
  \bibfield  {author} {\bibinfo {author} {\bibfnamefont {P.}~\bibnamefont
  {Mueller}} \emph {et~al.},\ }\href {\doibase 10.1103/PhysRevLett.99.252501}
  {\bibfield  {journal} {\bibinfo  {journal} {Phys. Rev. Lett.}\ }\textbf
  {\bibinfo {volume} {99}},\ \bibinfo {pages} {252501} (\bibinfo {year}
  {2007})}\BibitemShut {NoStop}%
\bibitem [{\citenamefont {Ewald}\ \emph {et~al.}(2004)\citenamefont {Ewald}
  \emph {et~al.}}]{Ewald2005}%
  \BibitemOpen
  \bibfield  {author} {\bibinfo {author} {\bibfnamefont {G.}~\bibnamefont
  {Ewald}} \emph {et~al.},\ }\href {\doibase 10.1103/PhysRevLett.93.113002}
  {\bibfield  {journal} {\bibinfo  {journal} {Phys. Rev. Lett.}\ }\textbf
  {\bibinfo {volume} {93}},\ \bibinfo {pages} {113002} (\bibinfo {year}
  {2004})}\BibitemShut {NoStop}%
\bibitem [{\citenamefont {S\'anchez}\ \emph {et~al.}(2006)\citenamefont
  {S\'anchez} \emph {et~al.}}]{Sanchez2006}%
  \BibitemOpen
  \bibfield  {author} {\bibinfo {author} {\bibfnamefont {R.}~\bibnamefont
  {S\'anchez}} \emph {et~al.},\ }\href {\doibase 10.1103/PhysRevLett.96.033002}
  {\bibfield  {journal} {\bibinfo  {journal} {Phys. Rev. Lett.}\ }\textbf
  {\bibinfo {volume} {96}},\ \bibinfo {pages} {033002} (\bibinfo {year}
  {2006})}\BibitemShut {NoStop}%
\bibitem [{\citenamefont {N\"ortersh\"auser}\ \emph {et~al.}(2009)\citenamefont
  {N\"ortersh\"auser} \emph {et~al.}}]{Noertershaeuser2009}%
  \BibitemOpen
  \bibfield  {author} {\bibinfo {author} {\bibfnamefont {W.}~\bibnamefont
  {N\"ortersh\"auser}} \emph {et~al.},\ }\href {\doibase
  10.1103/PhysRevLett.102.062503} {\bibfield  {journal} {\bibinfo  {journal}
  {Phys. Rev. Lett.}\ }\textbf {\bibinfo {volume} {102}},\ \bibinfo {pages}
  {062503} (\bibinfo {year} {2009})}\BibitemShut {NoStop}%
\bibitem [{\citenamefont {Krieger}\ \emph {et~al.}(2012)\citenamefont {Krieger}
  \emph {et~al.}}]{Krieger2012}%
  \BibitemOpen
  \bibfield  {author} {\bibinfo {author} {\bibfnamefont {A.}~\bibnamefont
  {Krieger}} \emph {et~al.},\ }\href {\doibase 10.1103/PhysRevLett.108.142501}
  {\bibfield  {journal} {\bibinfo  {journal} {Phys. Rev. Lett.}\ }\textbf
  {\bibinfo {volume} {108}},\ \bibinfo {pages} {142501} (\bibinfo {year}
  {2012})}\BibitemShut {NoStop}%
\bibitem [{\citenamefont {Yan}(2001)}]{Yan2001}%
  \BibitemOpen
  \bibfield  {author} {\bibinfo {author} {\bibfnamefont {Z.-C.}\ \bibnamefont
  {Yan}},\ }\href {\doibase 10.1103/PhysRevLett.86.5683} {\bibfield  {journal}
  {\bibinfo  {journal} {Phys. Rev. Lett.}\ }\textbf {\bibinfo {volume} {86}},\
  \bibinfo {pages} {5683} (\bibinfo {year} {2001})}\BibitemShut {NoStop}%
\bibitem [{\citenamefont {Yan}\ \emph {et~al.}(2008)\citenamefont {Yan},
  \citenamefont {N\"ortersh\"auser},\ and\ \citenamefont {Drake}}]{Yan2008}%
  \BibitemOpen
  \bibfield  {author} {\bibinfo {author} {\bibfnamefont {Z.-C.}\ \bibnamefont
  {Yan}}, \bibinfo {author} {\bibfnamefont {W.}~\bibnamefont
  {N\"ortersh\"auser}}, \ and\ \bibinfo {author} {\bibfnamefont {G.~W.~F.}\
  \bibnamefont {Drake}},\ }\href {\doibase 10.1103/PhysRevLett.100.243002}
  {\bibfield  {journal} {\bibinfo  {journal} {Phys. Rev. Lett.}\ }\textbf
  {\bibinfo {volume} {100}},\ \bibinfo {pages} {243002} (\bibinfo {year}
  {2008})}\BibitemShut {NoStop}%
\bibitem [{\citenamefont {Pachucki}\ and\ \citenamefont
  {Komasa}(2003)}]{Pachucki2003}%
  \BibitemOpen
  \bibfield  {author} {\bibinfo {author} {\bibfnamefont {K.}~\bibnamefont
  {Pachucki}}\ and\ \bibinfo {author} {\bibfnamefont {J.}~\bibnamefont
  {Komasa}},\ }\href {\doibase 10.1103/PhysRevA.68.042507} {\bibfield
  {journal} {\bibinfo  {journal} {Phys. Rev. A}\ }\textbf {\bibinfo {volume}
  {68}},\ \bibinfo {pages} {042507} (\bibinfo {year} {2003})}\BibitemShut
  {NoStop}%
\bibitem [{\citenamefont {Puchalski}\ and\ \citenamefont
  {Pachucki}(2008)}]{Puchalski2008}%
  \BibitemOpen
  \bibfield  {author} {\bibinfo {author} {\bibfnamefont {M.}~\bibnamefont
  {Puchalski}}\ and\ \bibinfo {author} {\bibfnamefont {K.}~\bibnamefont
  {Pachucki}},\ }\href {\doibase 10.1103/PhysRevA.78.052511} {\bibfield
  {journal} {\bibinfo  {journal} {Phys. Rev. A}\ }\textbf {\bibinfo {volume}
  {78}},\ \bibinfo {pages} {052511} (\bibinfo {year} {2008})}\BibitemShut
  {NoStop}%
\bibitem [{\citenamefont {Lu}\ \emph {et~al.}(2013)\citenamefont {Lu},
  \citenamefont {Mueller}, \citenamefont {Drake}, \citenamefont
  {N\"ortersh\"auser}, \citenamefont {Pieper},\ and\ \citenamefont
  {Yan}}]{Lu2013}%
  \BibitemOpen
  \bibfield  {author} {\bibinfo {author} {\bibfnamefont {Z.-T.}\ \bibnamefont
  {Lu}}, \bibinfo {author} {\bibfnamefont {P.}~\bibnamefont {Mueller}},
  \bibinfo {author} {\bibfnamefont {G.~W.~F.}\ \bibnamefont {Drake}}, \bibinfo
  {author} {\bibfnamefont {W.}~\bibnamefont {N\"ortersh\"auser}}, \bibinfo
  {author} {\bibfnamefont {S.~C.}\ \bibnamefont {Pieper}}, \ and\ \bibinfo
  {author} {\bibfnamefont {Z.-C.}\ \bibnamefont {Yan}},\ }\href {\doibase
  10.1103/RevModPhys.85.1383} {\bibfield  {journal} {\bibinfo  {journal} {Rev.
  Mod. Phys.}\ }\textbf {\bibinfo {volume} {85}},\ \bibinfo {pages} {1383}
  (\bibinfo {year} {2013})}\BibitemShut {NoStop}%
\bibitem [{\citenamefont {Puchalski}\ \emph {et~al.}(2015)\citenamefont
  {Puchalski}, \citenamefont {Komasa},\ and\ \citenamefont
  {Pachucki}}]{Puchalski2015}%
  \BibitemOpen
  \bibfield  {author} {\bibinfo {author} {\bibfnamefont {M.}~\bibnamefont
  {Puchalski}}, \bibinfo {author} {\bibfnamefont {J.}~\bibnamefont {Komasa}}, \
  and\ \bibinfo {author} {\bibfnamefont {K.}~\bibnamefont {Pachucki}},\ }\href
  {\doibase 10.1103/PhysRevA.92.062501} {\bibfield  {journal} {\bibinfo
  {journal} {Phys. Rev. A}\ }\textbf {\bibinfo {volume} {92}},\ \bibinfo
  {pages} {062501} (\bibinfo {year} {2015})}\BibitemShut {NoStop}%
\bibitem [{\citenamefont {Patk\'o\ifmmode~\check{s}\else \v{s}\fi{}}\ \emph
  {et~al.}(2016)\citenamefont {Patk\'o\ifmmode~\check{s}\else \v{s}\fi{}},
  \citenamefont {Yerokhin},\ and\ \citenamefont {Pachucki}}]{Patkos2016}%
  \BibitemOpen
  \bibfield  {author} {\bibinfo {author} {\bibfnamefont {V.~c.~v.}\
  \bibnamefont {Patk\'o\ifmmode~\check{s}\else \v{s}\fi{}}}, \bibinfo {author}
  {\bibfnamefont {V.~A.}\ \bibnamefont {Yerokhin}}, \ and\ \bibinfo {author}
  {\bibfnamefont {K.}~\bibnamefont {Pachucki}},\ }\href {\doibase
  10.1103/PhysRevA.94.052508} {\bibfield  {journal} {\bibinfo  {journal} {Phys.
  Rev. A}\ }\textbf {\bibinfo {volume} {94}},\ \bibinfo {pages} {052508}
  (\bibinfo {year} {2016})}\BibitemShut {NoStop}%
\bibitem [{\citenamefont {Patk\'o\ifmmode~\check{s}\else \v{s}\fi{}}\ \emph
  {et~al.}(2017)\citenamefont {Patk\'o\ifmmode~\check{s}\else \v{s}\fi{}},
  \citenamefont {Yerokhin},\ and\ \citenamefont {Pachucki}}]{Patkos2017}%
  \BibitemOpen
  \bibfield  {author} {\bibinfo {author} {\bibfnamefont {V.~c.~v.}\
  \bibnamefont {Patk\'o\ifmmode~\check{s}\else \v{s}\fi{}}}, \bibinfo {author}
  {\bibfnamefont {V.~A.}\ \bibnamefont {Yerokhin}}, \ and\ \bibinfo {author}
  {\bibfnamefont {K.}~\bibnamefont {Pachucki}},\ }\href {\doibase
  10.1103/PhysRevA.95.012508} {\bibfield  {journal} {\bibinfo  {journal} {Phys.
  Rev. A}\ }\textbf {\bibinfo {volume} {95}},\ \bibinfo {pages} {012508}
  (\bibinfo {year} {2017})}\BibitemShut {NoStop}%
\bibitem [{\citenamefont {Johansson}\ \emph {et~al.}(1993)\citenamefont
  {Johansson}, \citenamefont {Litzén}, \citenamefont {Kasten},\ and\
  \citenamefont {Kock}}]{Johansson1993}%
  \BibitemOpen
  \bibfield  {author} {\bibinfo {author} {\bibfnamefont {S.~G.}\ \bibnamefont
  {Johansson}}, \bibinfo {author} {\bibfnamefont {U.}~\bibnamefont {Litzén}},
  \bibinfo {author} {\bibfnamefont {J.}~\bibnamefont {Kasten}}, \ and\ \bibinfo
  {author} {\bibfnamefont {M.}~\bibnamefont {Kock}},\ }\href@noop {} {\bibfield
   {journal} {\bibinfo  {journal} {Astrophys. J.}\ }\textbf {\bibinfo {volume}
  {403}},\ \bibinfo {pages} {L25} (\bibinfo {year} {1993})}\BibitemShut
  {NoStop}%
\bibitem [{\citenamefont {N\"ortersh\"auser}\ \emph {et~al.}(2011)\citenamefont
  {N\"ortersh\"auser}, \citenamefont {Neff}, \citenamefont {S\'anchez},\ and\
  \citenamefont {Sick}}]{Noertershaeuser2011}%
  \BibitemOpen
  \bibfield  {author} {\bibinfo {author} {\bibfnamefont {W.}~\bibnamefont
  {N\"ortersh\"auser}}, \bibinfo {author} {\bibfnamefont {T.}~\bibnamefont
  {Neff}}, \bibinfo {author} {\bibfnamefont {R.}~\bibnamefont {S\'anchez}}, \
  and\ \bibinfo {author} {\bibfnamefont {I.}~\bibnamefont {Sick}},\ }\href
  {\doibase 10.1103/PhysRevC.84.024307} {\bibfield  {journal} {\bibinfo
  {journal} {Phys. Rev. C}\ }\textbf {\bibinfo {volume} {84}},\ \bibinfo
  {pages} {024307} (\bibinfo {year} {2011})}\BibitemShut {NoStop}%
\bibitem [{\citenamefont {Mrozowski}(1939)}]{Mrozowski1939}%
  \BibitemOpen
  \bibfield  {author} {\bibinfo {author} {\bibfnamefont {S.}~\bibnamefont
  {Mrozowski}},\ }\href {https://doi.org/10.1007/BF01340066} {\bibfield
  {journal} {\bibinfo  {journal} {Z. Physik}\ }\textbf {\bibinfo {volume}
  {112}} (\bibinfo {year} {1939})}\BibitemShut {NoStop}%
\bibitem [{\citenamefont {Burke}(1955)}]{Burke1955}%
  \BibitemOpen
  \bibfield  {author} {\bibinfo {author} {\bibfnamefont {E.~W.}\ \bibnamefont
  {Burke}},\ }\href {\doibase 10.1103/PhysRev.99.1839} {\bibfield  {journal}
  {\bibinfo  {journal} {Phys. Rev.}\ }\textbf {\bibinfo {volume} {99}},\
  \bibinfo {pages} {1839} (\bibinfo {year} {1955})}\BibitemShut {NoStop}%
\bibitem [{\citenamefont {Ekstr\"om}\ \emph {et~al.}(2015)\citenamefont
  {Ekstr\"om}, \citenamefont {Jansen}, \citenamefont {Wendt}, \citenamefont
  {Hagen}, \citenamefont {Papenbrock}, \citenamefont {Carlsson}, \citenamefont
  {Forss\'en}, \citenamefont {Hjorth-Jensen}, \citenamefont {Navr\'atil},\ and\
  \citenamefont {Nazarewicz}}]{Ekstroem2015}%
  \BibitemOpen
  \bibfield  {author} {\bibinfo {author} {\bibfnamefont {A.}~\bibnamefont
  {Ekstr\"om}}, \bibinfo {author} {\bibfnamefont {G.~R.}\ \bibnamefont
  {Jansen}}, \bibinfo {author} {\bibfnamefont {K.~A.}\ \bibnamefont {Wendt}},
  \bibinfo {author} {\bibfnamefont {G.}~\bibnamefont {Hagen}}, \bibinfo
  {author} {\bibfnamefont {T.}~\bibnamefont {Papenbrock}}, \bibinfo {author}
  {\bibfnamefont {B.~D.}\ \bibnamefont {Carlsson}}, \bibinfo {author}
  {\bibfnamefont {C.}~\bibnamefont {Forss\'en}}, \bibinfo {author}
  {\bibfnamefont {M.}~\bibnamefont {Hjorth-Jensen}}, \bibinfo {author}
  {\bibfnamefont {P.}~\bibnamefont {Navr\'atil}}, \ and\ \bibinfo {author}
  {\bibfnamefont {W.}~\bibnamefont {Nazarewicz}},\ }\href {\doibase
  10.1103/PhysRevC.91.051301} {\bibfield  {journal} {\bibinfo  {journal} {Phys.
  Rev. C}\ }\textbf {\bibinfo {volume} {91}},\ \bibinfo {pages} {051301}
  (\bibinfo {year} {2015})}\BibitemShut {NoStop}%
\bibitem [{\citenamefont {Entem}\ and\ \citenamefont
  {Machleidt}(2003)}]{Entem2003}%
  \BibitemOpen
  \bibfield  {author} {\bibinfo {author} {\bibfnamefont {D.~R.}\ \bibnamefont
  {Entem}}\ and\ \bibinfo {author} {\bibfnamefont {R.}~\bibnamefont
  {Machleidt}},\ }\href {\doibase 10.1103/PhysRevC.68.041001} {\bibfield
  {journal} {\bibinfo  {journal} {Phys. Rev. C}\ }\textbf {\bibinfo {volume}
  {68}},\ \bibinfo {pages} {041001} (\bibinfo {year} {2003})}\BibitemShut
  {NoStop}%
\bibitem [{\citenamefont {Roth}\ \emph {et~al.}(2012)\citenamefont {Roth},
  \citenamefont {Binder}, \citenamefont {Vobig}, \citenamefont {Calci},
  \citenamefont {Langhammer},\ and\ \citenamefont {Navr\'atil}}]{Roth2012}%
  \BibitemOpen
  \bibfield  {author} {\bibinfo {author} {\bibfnamefont {R.}~\bibnamefont
  {Roth}}, \bibinfo {author} {\bibfnamefont {S.}~\bibnamefont {Binder}},
  \bibinfo {author} {\bibfnamefont {K.}~\bibnamefont {Vobig}}, \bibinfo
  {author} {\bibfnamefont {A.}~\bibnamefont {Calci}}, \bibinfo {author}
  {\bibfnamefont {J.}~\bibnamefont {Langhammer}}, \ and\ \bibinfo {author}
  {\bibfnamefont {P.}~\bibnamefont {Navr\'atil}},\ }\href {\doibase
  10.1103/PhysRevLett.109.052501} {\bibfield  {journal} {\bibinfo  {journal}
  {Phys. Rev. Lett.}\ }\textbf {\bibinfo {volume} {109}},\ \bibinfo {pages}
  {052501} (\bibinfo {year} {2012})}\BibitemShut {NoStop}%
\bibitem [{\citenamefont {Hergert}\ \emph {et~al.}(2013)\citenamefont
  {Hergert}, \citenamefont {Binder}, \citenamefont {Calci}, \citenamefont
  {Langhammer},\ and\ \citenamefont {Roth}}]{Hergert2013}%
  \BibitemOpen
  \bibfield  {author} {\bibinfo {author} {\bibfnamefont {H.}~\bibnamefont
  {Hergert}}, \bibinfo {author} {\bibfnamefont {S.}~\bibnamefont {Binder}},
  \bibinfo {author} {\bibfnamefont {A.}~\bibnamefont {Calci}}, \bibinfo
  {author} {\bibfnamefont {J.}~\bibnamefont {Langhammer}}, \ and\ \bibinfo
  {author} {\bibfnamefont {R.}~\bibnamefont {Roth}},\ }\href {\doibase
  10.1103/PhysRevLett.110.242501} {\bibfield  {journal} {\bibinfo  {journal}
  {Phys. Rev. Lett.}\ }\textbf {\bibinfo {volume} {110}},\ \bibinfo {pages}
  {242501} (\bibinfo {year} {2013})}\BibitemShut {NoStop}%
\bibitem [{\citenamefont {Gebrerufael}\ \emph {et~al.}(2017)\citenamefont
  {Gebrerufael}, \citenamefont {Vobig}, \citenamefont {Hergert},\ and\
  \citenamefont {Roth}}]{Gebrerufael2017}%
  \BibitemOpen
  \bibfield  {author} {\bibinfo {author} {\bibfnamefont {E.}~\bibnamefont
  {Gebrerufael}}, \bibinfo {author} {\bibfnamefont {K.}~\bibnamefont {Vobig}},
  \bibinfo {author} {\bibfnamefont {H.}~\bibnamefont {Hergert}}, \ and\
  \bibinfo {author} {\bibfnamefont {R.}~\bibnamefont {Roth}},\ }\href {\doibase
  10.1103/PhysRevLett.118.152503} {\bibfield  {journal} {\bibinfo  {journal}
  {Phys. Rev. Lett.}\ }\textbf {\bibinfo {volume} {118}},\ \bibinfo {pages}
  {152503} (\bibinfo {year} {2017})}\BibitemShut {NoStop}%
\bibitem [{\citenamefont {Entem}\ \emph {et~al.}(2017)\citenamefont {Entem},
  \citenamefont {Machleidt},\ and\ \citenamefont {Nosyk}}]{Entem2017}%
  \BibitemOpen
  \bibfield  {author} {\bibinfo {author} {\bibfnamefont {D.~R.}\ \bibnamefont
  {Entem}}, \bibinfo {author} {\bibfnamefont {R.}~\bibnamefont {Machleidt}}, \
  and\ \bibinfo {author} {\bibfnamefont {Y.}~\bibnamefont {Nosyk}},\ }\href
  {\doibase 10.1103/PhysRevC.96.024004} {\bibfield  {journal} {\bibinfo
  {journal} {Phys. Rev. C}\ }\textbf {\bibinfo {volume} {96}},\ \bibinfo
  {pages} {024004} (\bibinfo {year} {2017})}\BibitemShut {NoStop}%
\bibitem [{\citenamefont {H\"uther}\ \emph {et~al.}()\citenamefont {H\"uther},
  \citenamefont {Vobig},\ and\ \citenamefont {Roth}}]{Huether}%
  \BibitemOpen
  \bibfield  {author} {\bibinfo {author} {\bibfnamefont {T.}~\bibnamefont
  {H\"uther}}, \bibinfo {author} {\bibfnamefont {K.}~\bibnamefont {Vobig}}, \
  and\ \bibinfo {author} {\bibfnamefont {R.}~\bibnamefont {Roth}},\ }\href@noop
  {} {}\bibinfo {note} {(in preparation)}\BibitemShut {NoStop}%
\bibitem [{\citenamefont {Carlson}\ \emph {et~al.}(2015)\citenamefont
  {Carlson}, \citenamefont {Gandolfi}, \citenamefont {Pederiva}, \citenamefont
  {Pieper}, \citenamefont {Schiavilla}, \citenamefont {Schmidt},\ and\
  \citenamefont {Wiringa}}]{Carlson2015}%
  \BibitemOpen
  \bibfield  {author} {\bibinfo {author} {\bibfnamefont {J.}~\bibnamefont
  {Carlson}}, \bibinfo {author} {\bibfnamefont {S.}~\bibnamefont {Gandolfi}},
  \bibinfo {author} {\bibfnamefont {F.}~\bibnamefont {Pederiva}}, \bibinfo
  {author} {\bibfnamefont {S.~C.}\ \bibnamefont {Pieper}}, \bibinfo {author}
  {\bibfnamefont {R.}~\bibnamefont {Schiavilla}}, \bibinfo {author}
  {\bibfnamefont {K.~E.}\ \bibnamefont {Schmidt}}, \ and\ \bibinfo {author}
  {\bibfnamefont {R.~B.}\ \bibnamefont {Wiringa}},\ }\href {\doibase
  10.1103/RevModPhys.87.1067} {\bibfield  {journal} {\bibinfo  {journal} {Rev.
  Mod. Phys.}\ }\textbf {\bibinfo {volume} {87}},\ \bibinfo {pages} {1067}
  (\bibinfo {year} {2015})}\BibitemShut {NoStop}%
\bibitem [{\citenamefont {Wiringa}\ \emph {et~al.}(1995)\citenamefont
  {Wiringa}, \citenamefont {Stoks},\ and\ \citenamefont
  {Schiavilla}}]{Wiringa1995}%
  \BibitemOpen
  \bibfield  {author} {\bibinfo {author} {\bibfnamefont {R.~B.}\ \bibnamefont
  {Wiringa}}, \bibinfo {author} {\bibfnamefont {V.~G.~J.}\ \bibnamefont
  {Stoks}}, \ and\ \bibinfo {author} {\bibfnamefont {R.}~\bibnamefont
  {Schiavilla}},\ }\href {\doibase 10.1103/PhysRevC.51.38} {\bibfield
  {journal} {\bibinfo  {journal} {Phys. Rev.}\ }\textbf {\bibinfo {volume}
  {C51}},\ \bibinfo {pages} {38} (\bibinfo {year} {1995})}\BibitemShut
  {NoStop}%
\bibitem [{\citenamefont {Pieper}\ \emph {et~al.}(2001)\citenamefont {Pieper},
  \citenamefont {Pandharipande}, \citenamefont {Wiringa},\ and\ \citenamefont
  {Carlson}}]{Pieper2001}%
  \BibitemOpen
  \bibfield  {author} {\bibinfo {author} {\bibfnamefont {S.~C.}\ \bibnamefont
  {Pieper}}, \bibinfo {author} {\bibfnamefont {V.~R.}\ \bibnamefont
  {Pandharipande}}, \bibinfo {author} {\bibfnamefont {R.~B.}\ \bibnamefont
  {Wiringa}}, \ and\ \bibinfo {author} {\bibfnamefont {J.}~\bibnamefont
  {Carlson}},\ }\href {\doibase 10.1103/PhysRevC.64.014001} {\bibfield
  {journal} {\bibinfo  {journal} {Phys. Rev.}\ }\textbf {\bibinfo {volume}
  {C64}},\ \bibinfo {pages} {014001} (\bibinfo {year} {2001})}\BibitemShut
  {NoStop}%
\bibitem [{\citenamefont {Pieper}(2008)}]{Pieper2008}%
  \BibitemOpen
  \bibfield  {author} {\bibinfo {author} {\bibfnamefont {S.~C.}\ \bibnamefont
  {Pieper}},\ }\href {\doibase 10.1063/1.2932280} {\bibfield  {journal}
  {\bibinfo  {journal} {AIP Conf. Proc.}\ }\textbf {\bibinfo {volume} {1011}},\
  \bibinfo {pages} {143} (\bibinfo {year} {2008})}\BibitemShut {NoStop}%
\bibitem [{\citenamefont {Lonardoni}\ \emph {et~al.}(2018)\citenamefont
  {Lonardoni}, \citenamefont {Gandolfi}, \citenamefont {Lynn}, \citenamefont
  {Petrie}, \citenamefont {Carlson}, \citenamefont {Schmidt},\ and\
  \citenamefont {Schwenk}}]{Lonardoni2018}%
  \BibitemOpen
  \bibfield  {author} {\bibinfo {author} {\bibfnamefont {D.}~\bibnamefont
  {Lonardoni}}, \bibinfo {author} {\bibfnamefont {S.}~\bibnamefont {Gandolfi}},
  \bibinfo {author} {\bibfnamefont {J.~E.}\ \bibnamefont {Lynn}}, \bibinfo
  {author} {\bibfnamefont {C.}~\bibnamefont {Petrie}}, \bibinfo {author}
  {\bibfnamefont {J.}~\bibnamefont {Carlson}}, \bibinfo {author} {\bibfnamefont
  {K.~E.}\ \bibnamefont {Schmidt}}, \ and\ \bibinfo {author} {\bibfnamefont
  {A.}~\bibnamefont {Schwenk}},\ }\href {\doibase 10.1103/PhysRevC.97.044318}
  {\bibfield  {journal} {\bibinfo  {journal} {Phys. Rev.}\ }\textbf {\bibinfo
  {volume} {C97}},\ \bibinfo {pages} {044318} (\bibinfo {year}
  {2018})}\BibitemShut {NoStop}%
\bibitem [{\citenamefont {Lovato}\ \emph {et~al.}(2013)\citenamefont {Lovato},
  \citenamefont {Gandolfi}, \citenamefont {Butler}, \citenamefont {Carlson},
  \citenamefont {Lusk}, \citenamefont {Pieper},\ and\ \citenamefont
  {Schiavilla}}]{Lovato2013}%
  \BibitemOpen
  \bibfield  {author} {\bibinfo {author} {\bibfnamefont {A.}~\bibnamefont
  {Lovato}}, \bibinfo {author} {\bibfnamefont {S.}~\bibnamefont {Gandolfi}},
  \bibinfo {author} {\bibfnamefont {R.}~\bibnamefont {Butler}}, \bibinfo
  {author} {\bibfnamefont {J.}~\bibnamefont {Carlson}}, \bibinfo {author}
  {\bibfnamefont {E.}~\bibnamefont {Lusk}}, \bibinfo {author} {\bibfnamefont
  {S.~C.}\ \bibnamefont {Pieper}}, \ and\ \bibinfo {author} {\bibfnamefont
  {R.}~\bibnamefont {Schiavilla}},\ }\href {\doibase
  10.1103/PhysRevLett.111.092501} {\bibfield  {journal} {\bibinfo  {journal}
  {Phys. Rev. Lett.}\ }\textbf {\bibinfo {volume} {111}},\ \bibinfo {pages}
  {092501} (\bibinfo {year} {2013})}\BibitemShut {NoStop}%
\bibitem [{\citenamefont {{Stovall}}\ \emph {et~al.}(1966)\citenamefont
  {{Stovall}}, \citenamefont {{Goldemberg}},\ and\ \citenamefont
  {{Isabelle}}}]{Stovall1966}%
  \BibitemOpen
  \bibfield  {author} {\bibinfo {author} {\bibfnamefont {T.}~\bibnamefont
  {{Stovall}}}, \bibinfo {author} {\bibfnamefont {J.}~\bibnamefont
  {{Goldemberg}}}, \ and\ \bibinfo {author} {\bibfnamefont {D.}~\bibnamefont
  {{Isabelle}}},\ }\href {\doibase 10.1016/0029-5582(66)90302-6} {\bibfield
  {journal} {\bibinfo  {journal} {Nucl. Phys.}\ }\textbf {\bibinfo {volume}
  {86}},\ \bibinfo {pages} {225} (\bibinfo {year} {1966})}\BibitemShut
  {NoStop}%
\end{thebibliography}%
\end{document}